\newcommand\etal{\it et al. \rm}
\newcommand\hmpc{h^{-1}{\rm Mpc}}
\newcommand\hkpc{h^{-1}{\rm kpc}}
\newcommand\hmsun{h^{-1}M_{\sun}}
\newcommand\msun{M_{\sun}}
\newcommand\gsim{ \lower .75ex \hbox{$\sim$} \llap{\raise .27ex \hbox{$>$}} }
\newcommand\lsim{ \lower .75ex \hbox{$\sim$} \llap{\raise .27ex \hbox{$<$}} }

\newcommand\LCDM{$\Lambda{\rm CDM}$}
\newcommand\tCDM{$\tau{\rm CDM}$}

\newcommand\OCDM{OCDM}

\newcommand\pthreem{${\rm P}^3{\rm M}$}
\input epsf.sty

\documentstyle[11pt,paspconf]{article}

\begin{document}
\title{The Virgo consortium: simulations of dark matter and galaxy
clustering} 
\author{A. Jenkins  and C. S. Frenk}
\affil{Physics Dept, University of Durham, Durham, DH1 3LE}
\author{F.R. Pearce, P.A. Thomas, R. Hutchings}
\affil{MAPS, University of Sussex, Brighton BN1 9QH}
\author{J. M. Colberg, S. D. M. White}
\affil{Max-Planck Institute for Astrophysics, Garching, Munich, D-85740}
\author{H. M. P. Couchman}
\affil{ Department of Astronomy, University of Western 
Ontario, London, Ontario N6A 3K7, Canada}
\author{J. A. Peacock}
\affil{ Royal Observatory, Blackford Hill, Edinburgh, EH9 3HJ} 
\author{G. P. Efstathiou}
\affil{Dept of Physics, Nuclear Physics Building, Keble Road,
Oxford, OX1 3RH}
\author{A. H. Nelson}
\affil{ Dept of Physics and Astronomy, UWCC, 
PO Box 913, Cardiff}



\begin{abstract}
We report on work in progress by the Virgo consortium, a collaboration set
up to carry out large simulations of the formation of galaxies and
large-scale structure exploiting the latest generation of parallel
supercomputers. We show results of $256^3$ particle N-body simulations of
the clustering evolution of dark matter in four cold dark matter models
with different cosmological parameters. The high resolution and large
volume of these simulations allows us to determine reliably the mass
autocorrelation function for pair separations in the range $40\hkpc$ to
$20\hmpc$. Comparison of these with the observed galaxy correlation
function shows that for any of these models to be viable, the distribution
of galaxies must be biased relative to the distribution of mass in a
non-trivial, scale-dependent fashion. In particular, low $\Omega_0$ models
require the galaxies to be more {\it weakly} clustered than the mass at
small and intermediate pair separations. Simulations which include the
evolution of gas show that cold gas knots form with approximately the
abundance expected on theoretical grounds, although a few excessively
massive objects grow near the centres of rich clusters.  The locations
where these cold gas knots form are, in general, biased relative to the
distribution of mass in a scale-dependent way. Some of these biases have
the required sign but they are, for the most part, weaker than is necessary
for agreement with observations. The antibias present in our low $\Omega_0$
N-body/SPH simulation appears to be related to the merging and disruption
of galaxies in rich clusters.

\end{abstract}




\section{Introduction}
 
Understanding the relationship between the distribution of dark matter and
the distribution of visible galaxies is one of the central problems of
modern cosmology. Theoretically, the problem of calculating the statistical
properties of the mass distribution in a given cosmological model (and here
we will consider only hierarchical clustering models) consists primarily of
understanding the gravitational dynamics of collisionless dark matter,
as the baryonic component contributes only a small fraction of the total
mass density. The gravitational evolution of dark matter universes has
been extensively investigated using N-body simulations (for a review 
see Frenk
1991) and analytic approximations such as the Press \& Schechter (1974)
formalism and its extensions (Bower 1992, Bond \etal 1991) or second order 
perturbation theory (Bouchet \etal\ (1992), Bernardeau (1994)). 

With the exception of gravitational lensing studies, a subject still in its
infancy, the only information we have on the properties of the underlying
mass field, comes indirectly through studies of the baryonic component of the
Universe, primarily galaxies, but also X-ray emitting gas in clusters and
groups. Our lack of understanding of how the properties of the matter we
can observe are related to the properties of the underlying mass field is a
major obstacle in assessing the validity of theoretical models of structure
formation. With the simplest possible assumption, that light traces mass
(at least above some scale larger than that of galaxies where baryons
dominate), hierarchical clustering theories do lead to model universes
which superficially resemble are own. There are, however, many
disagreements of detail with observations and there is much debate on how
to remove these either by a suitable choice of cosmological parameters, by
varying the assumptions about the nature of the dark matter or by
arguing that the observed baryons are biased tracers of the mass.

A way out of this impasse is to try and predict the behaviour of baryonic
matter from first principles. Unfortunately, the dissipative physics that
govern the evolution of gas and the formation of stars makes this an
exceedingly complex problem. One approach is to describe the behaviour of
baryons within a merging hierarchy of dark matter halos using 
simple rules based on analytic or heuristic considerations. Such
semianalytic modelling has proved quite successful in explaining some of
the observed properties of galaxies ({\it e.g.} White and Frenk 1991; Kauffman
\etal 1993; Cole \etal 1994). The more direct approach, which is allied to
N-body simulations of the dark matter, is to try and calculate numerically
the hydrodynamic evolution of baryons, using heuristic rules for describing
processes associated with star formation. This approach requires formidable
computing power which is only now becoming available (Cen \& Ostriker 1996,
Katz \etal 1992, Evrard, Summers and Davis 1994, Frenk \etal 1996)

The Virgo consortium was formed with the aim of exploting the latest
generation of parallel supercomputers in order to address the problem of
calculating the coupled evolution of dark matter and gas in the expanding
universe. This kind of work represents a relatively new frontier where the
very choice of numerical techniques and input physics is still very much an
issue. Ultimately, it may prove possible to make reliable and testable
predictions of, for example, the abundance and clustering pattern of
galaxies expected in specific cosmologies. Here we give a short progress
report on our programme. In section 2 we provide a very brief description
of our numerical code and its current input physics. In section 3 we
discuss results from large N-body simulations which reveal the kind of
relationship required between the clustering of galaxies and dark
matter in a selection of popular cosmological models. In section 4 we
present preliminary results of two large N-body/gasdynamic simulations
which begin to address the extent to which these requirements are met in
practice.  This work is ongoing and we refrain from drawing any strong
conclusions at present.

\section{The Code}

Our current working code is based on ``Hydra,'' a code developed by
Couchman, Thomas and Pearce (1995) and parallelized by Pearce \etal
(1995). Gravity is treated using the adaptive \pthreem\ technique and
gas dynamics are treated using ``Smooth Particle Hydrodynamics''(SPH). The smallest
mesh refinements, placed on the regions of strongest clustering, are farmed
out to individual processors. Larger refinements and the base level grid
are run in parallel across all the processors.  For a typical dark matter
only run with $256^3$ particles, we use a $512^3$ grid as the base mesh and
the code is run on 256 processors.
\begin{figure}
\plotone{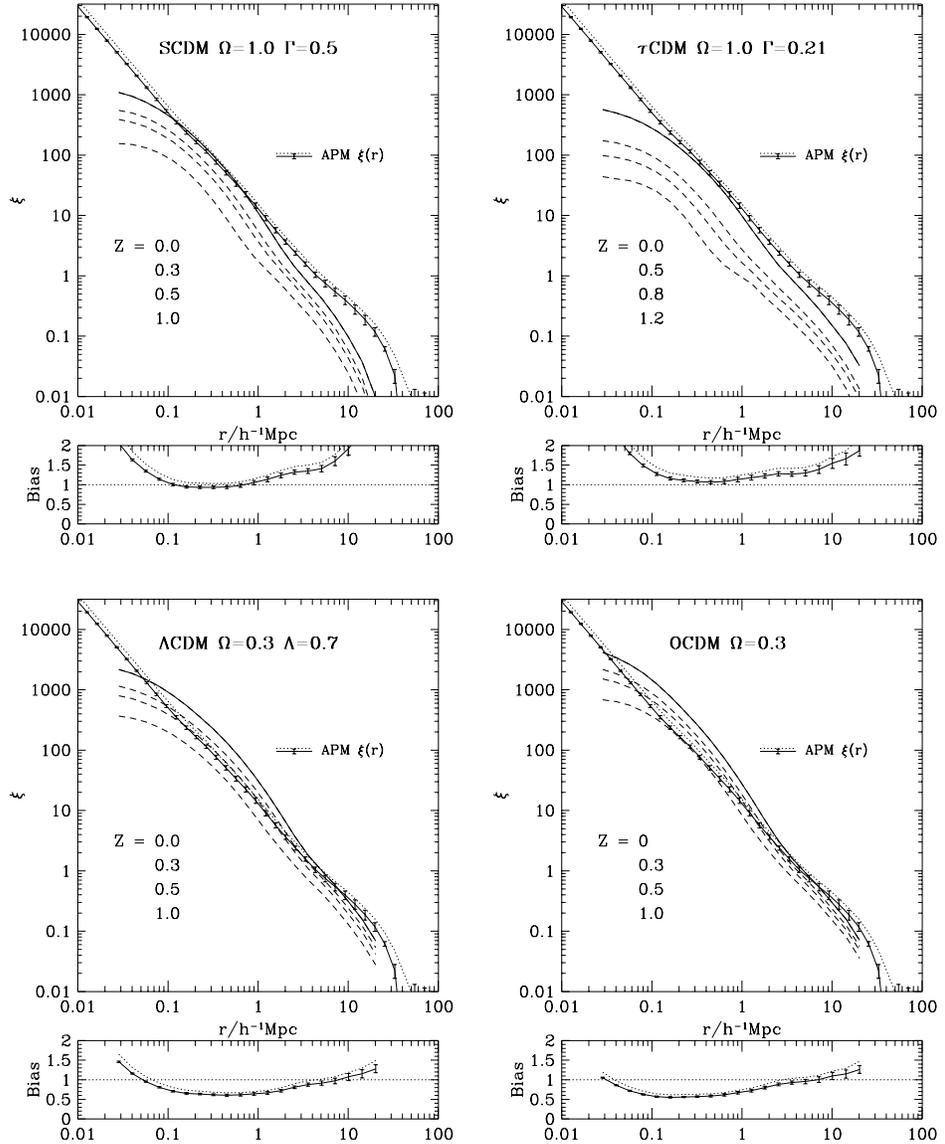}
\caption{ The mass autocorrelation function in 4 different cosmologies. The
dashed lines show $\xi(r)$ at different redshifts as indicated in the
figure legend. Baugh's (1996) estimates of the observed galaxy correlation
function, based on the APM galaxy survey, are also shown: the solid curve
assumes no evolution in the comoving clustering pattern, while the dotted
curve assumes linear evolution on all scales. Error bars are shown only for
the first estimate but they are similar for both. The lower plot in each
panel gives the square root of the ratio of the observed galaxy correlation
function to the predicted mass correlation function at z=0. This is the
scale dependent bias required for each model to match observations.}
\end{figure}

Although our code can treat the effects of a photoionizing background and,
through a simple heuristic model, the effects of star formation and
feedback, the gas simulations described below follow only the evolution of
gas subject to radiative cooling and shocks. All our simulations have been
carried out on Cray-T3Ds at the Edinburgh Parallel Supercomputer Centre
and at the Max-Planck Rechen Zentrum in Garching. 

\section {Dark Matter Simulations}

We have carried out a set of very large N-body simulations of cold dark
matter universes with four different choices of cosmological
parameters. With the exception of a $200^3$ particle open model, each
simulation followed the evolution of $256^3$ particles in a computational
box of $239.5\hmpc$ on side. (Here and below $h$ denotes Hubble's constant
in units of 100 km/s/Mpc). The gravitational softening was typically
$20-30\hkpc$. In all cases, the initial fluctuation amplitude was set by
requiring that the model should reproduce the observed abundance of rich clusters
at the present day. This was accomplished by setting, $\sigma_8$, the
present day linear rms fluctuation on spheres of radius $8\hmpc$, to the
values recommended by White, Efstathiou \& Frenk (1993) and Eke, Cole \&
Frenk (1996).

Figure~1 shows mass autocorrelation functions, $\xi(r)$, at several epochs
in all four simulations. Proceeding anticlockwise from the top left panel,
the models plotted are: ``standard'' CDM (SCDM with $\Omega=1$, $h=0.5$,
$\sigma_8=0.61$); ``Lambda'' CDM (\LCDM\ with $\Omega_0 = 0.3$, $\Lambda_0
= 0.7$, $h=0.7$, $\sigma_8=0.9$); ``Open'' CDM (\OCDM\ with $\Omega_0=0.3$,
$h=0.21$, $\sigma_8=0.85$); and ``Tau'' CDM (\tCDM\ with $\Omega= 1.0$,
$h=0.5$, $\sigma_8=0.61$ and the same power spectrum shape as \LCDM\ and
\OCDM). Because of the high resolution and large volume of 
these simulations the autocorrelation function is reliably determined over
a large range of scales. The shape of $\xi(r)$ is broadly similar in all
the models: it is relatively shallow on small scales, becomes increasingly
steep and has an inflection point at $\xi\simeq 1$. The SCDM case shows the
steepest drop at pair separations $\gsim 10\hmpc$.

Also plotted in Figure~1 are estimates of the autocorrelation function
of galaxies as inferred by Baugh (1996) from the angular correlation
function of the APM galaxy survey (Maddox \etal 1990) and a model for
the evolution of $\xi_{gal}(r)$. The solid curve with error bars shows
the estimate assuming no evolution of clustering in comoving
coordinates while the dotted curve shows the estimate assuming linear
evolution of clustering on all scales. (The statistical errors are
similar in both cases). Except in the
\OCDM\ case, $\xi(r)$ always falls below the galaxy data on very 
small scales. In the two $\Omega=1$ models, the mass correlation
function matches the galaxy data out to pair separations $\sim 1\hmpc$
and then falls below them. In the low $\Omega_0$ cases, on the other
hand, $\xi(r)$ rises above the galaxy data out to pair separations of
a few $\hmpc$ and then closely follows $\xi_{gal}(r)$ before falling
slightly below it at the largest pair separations. The lower box in
each panel shows the square root of the ratio of the galaxy
correlation function to the mass correlation function. This is the
bias in the galaxy distribution that would be required for each model
to match the observed strength of galaxy clustering. In all cases, the
bias function is scale dependent. In the $\Omega=1$ models, it is
approximately unity at small and intermediate pair separation before
rising steeply at large separations (particularly for the SCDM
model). In our low $\Omega_0$ models, the bias falls below unity over
a large range of pair separations indicating the need for a
substantial negative bias or antibias.

Our analysis demonstrates that for the models under discussion to be viable
a non-trivial relationship must exist between the distributions of galaxies
and mass. These dark matter only simulations cannot be used to decide if
such biases are plausible since this requires modelling the process of
galaxy formation. In the next section we discuss our first attempts to
address this problem.

\section{Gas Simulations}

We have carried out two N-body/SPH simulations that follow the coupled
evolution of gas and dark matter in the SCDM and \LCDM\ cosmologies
discussed in the preceeding section. (Because of a programming error the
value of $\sigma_8$ in the \LCDM\ model was set to 1.4 instead of the
desired value of 0.9). Each species is represented by 2 million
particles. The gas is allowed to cool radiatively according to the cooling
function for a fully ionized plasma with half-solar 
metallicity. Experimentation shows that efficient cooling on galactic
scales requires a gas particle mass of no more than $2\times10^9\msun$.  We
adopt the value of the baryon density, $\Omega_b$, implied by
nucleosynthesis considerations: $\Omega_bh^2 = 0.015$ (Copi \etal
1995). With this choice of parameters the computational volume is 100 Mpc
on a side in both models.
\begin{figure}
\plottwo{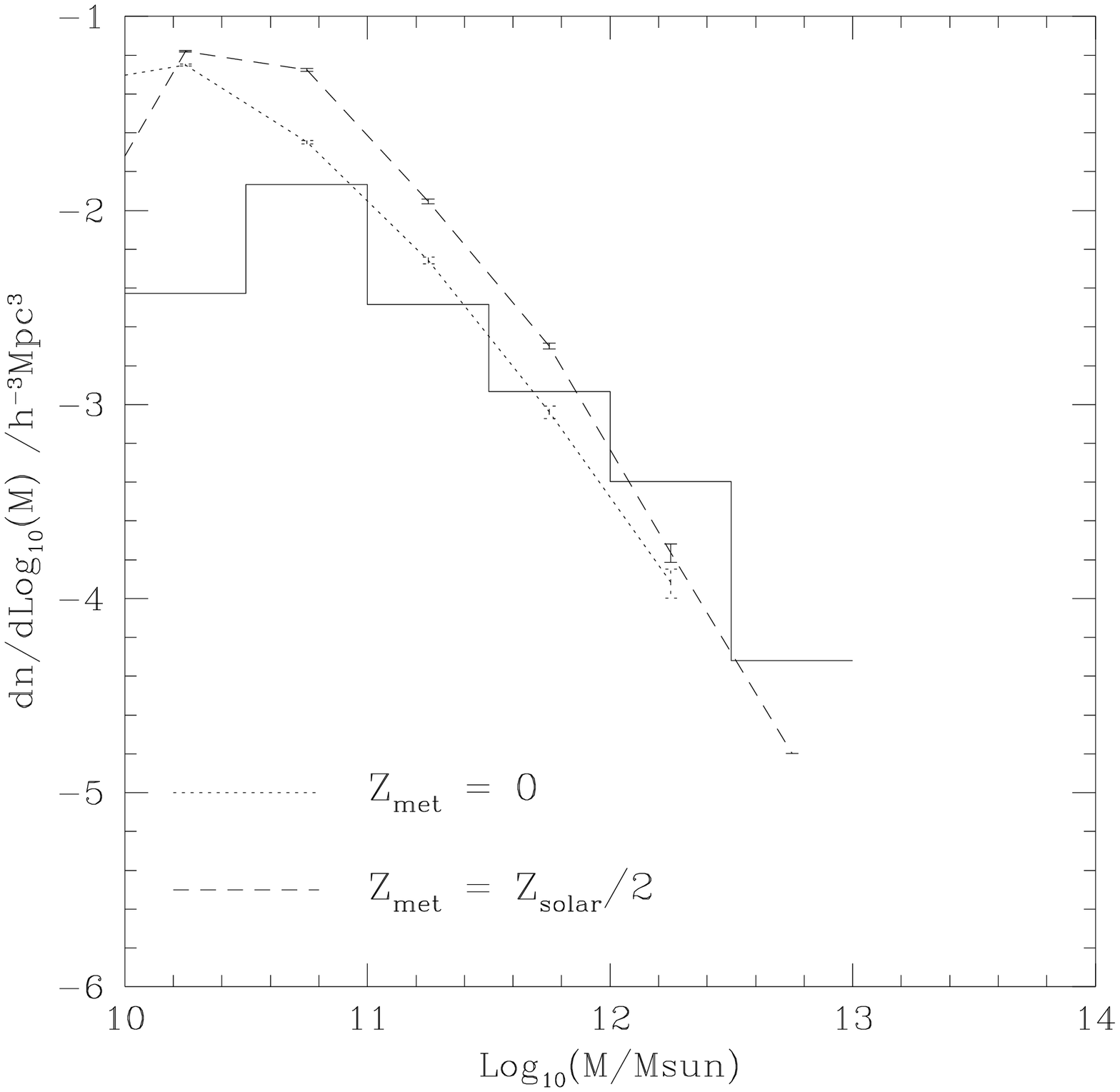}{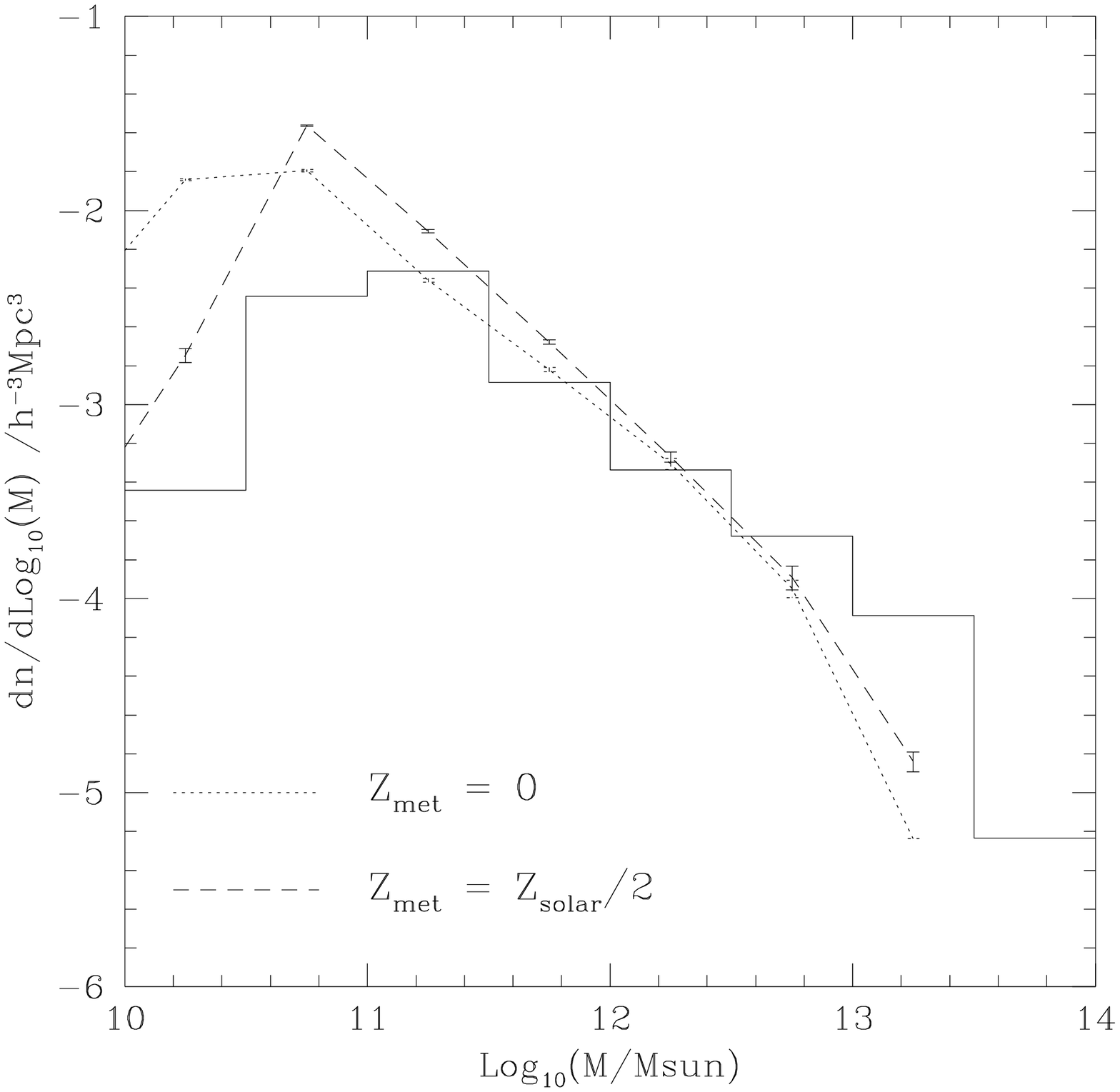}
\caption{Baryonic mass function of galaxies in 
two N-body/SPH simulations: ``standard'' CDM (left) and ``Lambda''
($\Omega_0 = 0.3$, $\Lambda_0 = 0.7$) CDM (right).  The simulation results
are plotted as histograms. The curves show predictions for the cold gas
mass function from a suitably modified version of the galaxy formation
semianalytic model of Cole \etal (1994; Baugh, private
communication). Results are given for two different metallicities,
primordial (dotted line) and half-solar (dashed line). The simulations
assume half-solar gas metallicity.}
\end{figure}

The simulations start at redshift $z=25$ in the SCDM model and $z=50$ in
the \LCDM\ model.  Structure in the dark matter builds up in the familiar
hierarchical fashion. The gas is able to cool radiatively into the evolving
population of dark matter halos. There it settles into cold knots of size
comparable to the gravitational softening length, $\sim 20$ kpc. These cold
knots we identify with ``galaxies.'' At the present epoch, the number of
galaxies with more than 20 gas particles is 1200 for SCDM and 1800 for
\LCDM. In Figure~2 we plot the baryonic mass function in the two
models. The number densities decline at the low mass end due to resolution
effects. Above the resolution limit, the number densities fall off with 
mass approximately as a power law, with perhaps an indication of a high mass
cutoff.

One way to check whether these mass functions are correct, {\it ie} whether
they are the expected outcome of the input physics, is to compare them with
the results of the semianalytic models mentioned in Section~1. These can be
tailored to include the same physics as the simulations, a 
mass resolution limit and gas subject only to shock heating and radiative
cooling. The mass functions predicted by the appropriately modified version
of the semianalytic model of Cole \etal (1994) are plotted in Figure~2 (Baugh,
private communication). A mass resolution corresponding to 20 SPH particles
has been assumed and results are shown for two gas metallicities, primordial
(dotted line) and half-solar (dashed line). At the low mass end, the
semianalytic calculation predicts many more ``galaxies'' than formed in the
simulations. This indicates that the effective mass resolution of the
simulations is closer to 100 SPH particle masses than to 20. Indeed, beyond
about $10^{11}\msun$, the agreement between the numerical and semianalytic
results is good, particularly in view of the fact that there are no adjustable
parameters in this comparison. At the high mass end both simulations, but
particularly the \LCDM\ model, produce too many objects. These very massive
``galaxies'' occur predominantly in the cores of rich clusters. Thus the
simulations seem to suffer from an ``overmerging'' problem similar to that
seen in the simulations of Frenk \etal (1996). This problem affects only a
small fraction of the mass: the total amount of cold gas in the models and
in the semianalytic models is very similar.

\begin{figure}
\plottwo{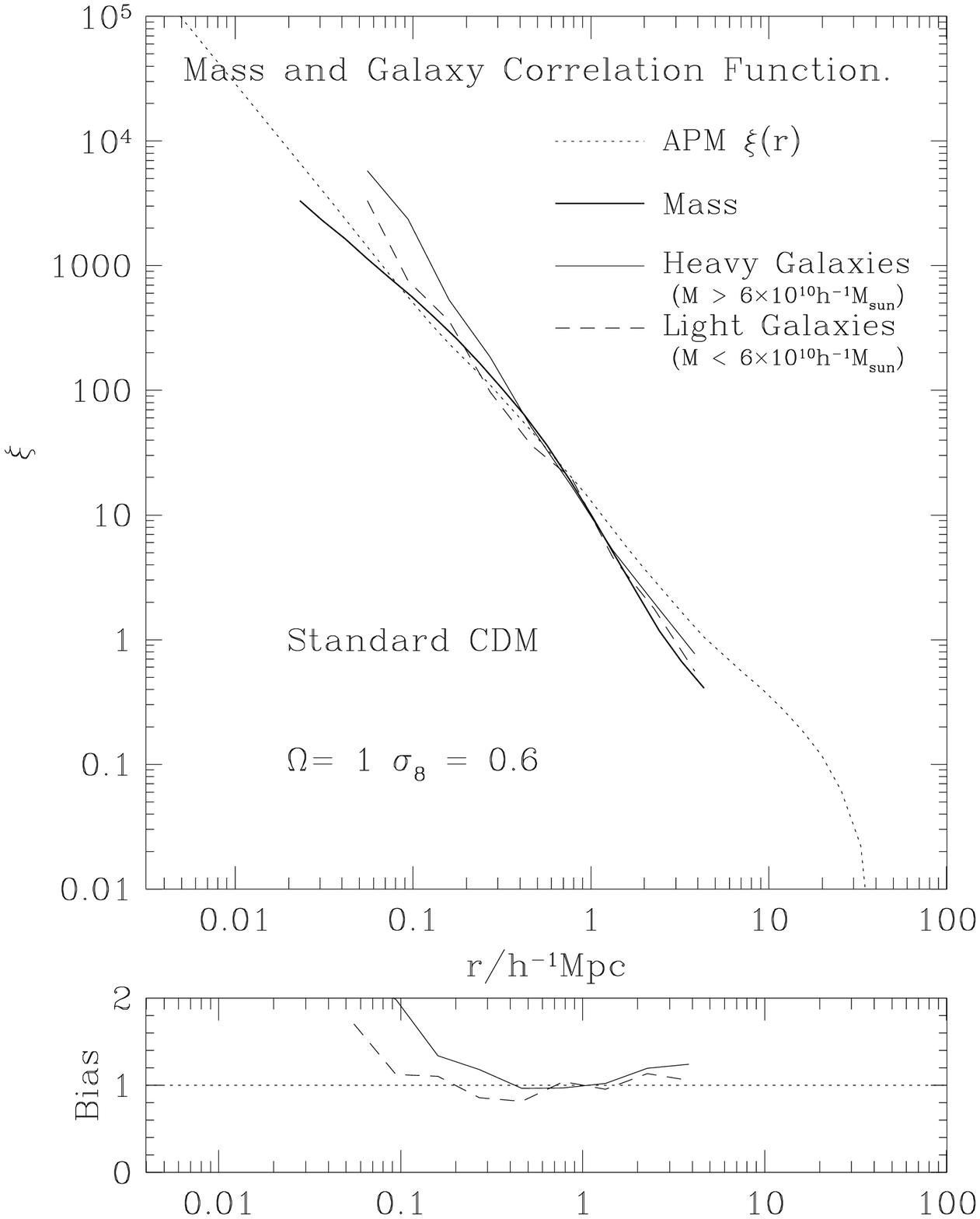}{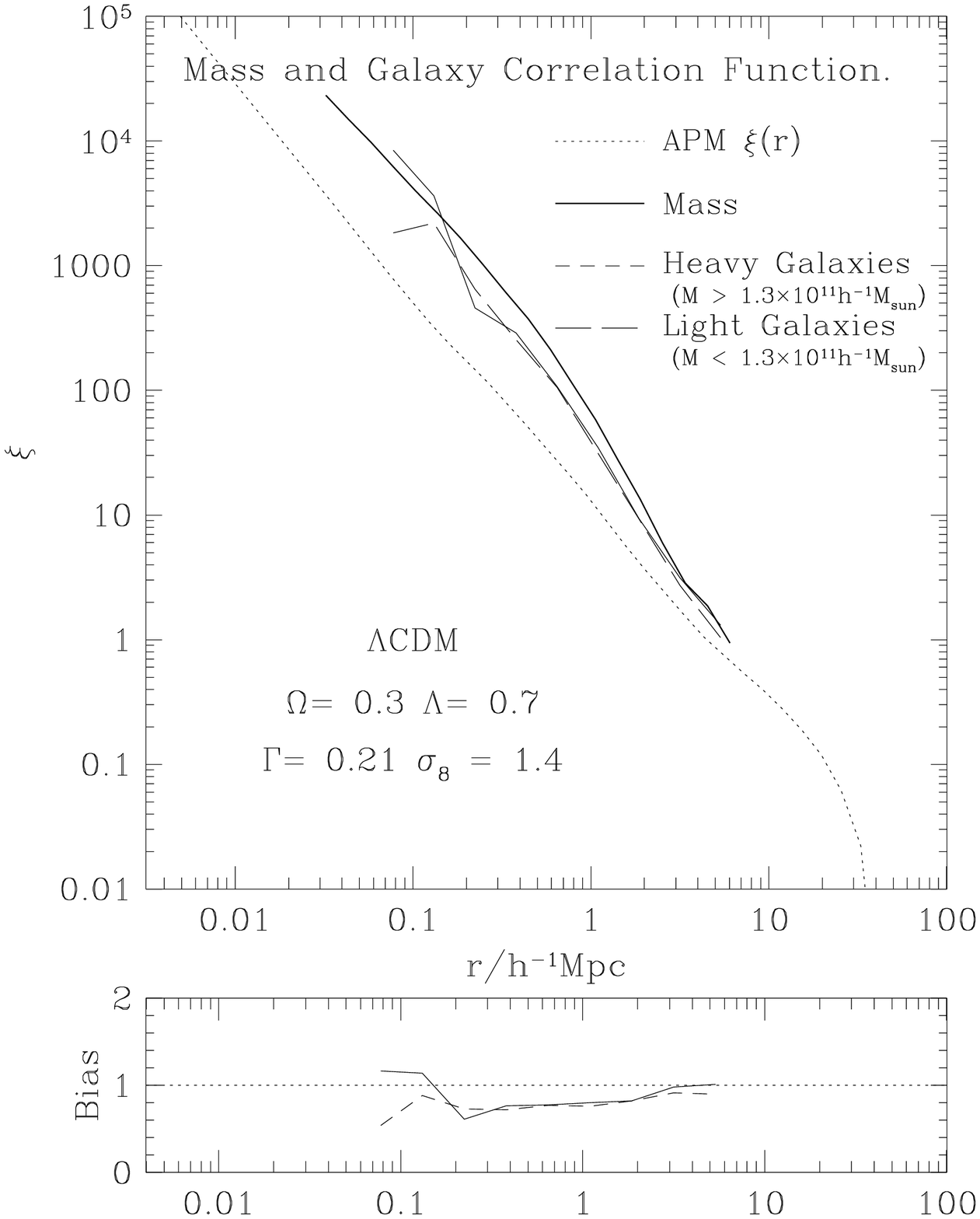}
\caption{Autocorrelation functions for dark matter and galaxies in two 
N-body/SPH simulations. The dotted line in the larger box shows the
observed galaxy correlation function (assuming no evolution 
-- see the caption to Figure~1) The heavy curve shows the mass
correlation function and the other two curves show the correlation
functions for galaxies in two different mass bins as indicated in the
legend.  The lower plot in each panel gives the square root of ratio of the
model galaxy correlation function to the correlation function of the mass.}

\end{figure}

Autocorrelation functions for both galaxies and mass in our two simulations
are shown in Figure~3. We split the galaxy samples into two halves,
``heavy'' and ``light'' galaxies with a cutoff mass of $M_{gas}= 6\times
10^{10}\hmsun$ in the SCDM model and $M_{gas}= 1.3\times 10^{11}\hmsun$ in
the \LCDM\ model. Apart from a moderate effect in the SCDM case at
separations $\lsim 1\hmpc$, there is little mass segregation when the
sample is split in this way. The galaxy correlation functions are compared
with the mass correlation function in the ``bias'' plots shown in the lower
part of each panel. In the SCDM case the correlation functions of both
heavy and light galaxies are positively biased relative to the mass on
scales less than a few hundred $\hkpc$ and larger than about $2\hmpc$. By
contrast, in the \LCDM\ case, galaxies of both types are antibiased over
most of the distance range considered but they closely trace the mass at
separations $\gsim 3 \hmpc$.  The observed galaxy correlation function,
reproduced from Figure~1, is shown as a dotted line in Figure~3. (For
clarity we plot only the ``no evolution'' estimate).  Although the various
bias effects seen in the simulations seem to have the correct sign, they
are, for the most part, weaker than required to account for the
observations. We note, however, that our adopted value of $\sigma_8$ in the
\LCDM\ model was erroneously set too high and this might well have weakened
the biases present in this model.

\begin{figure}
\plotone{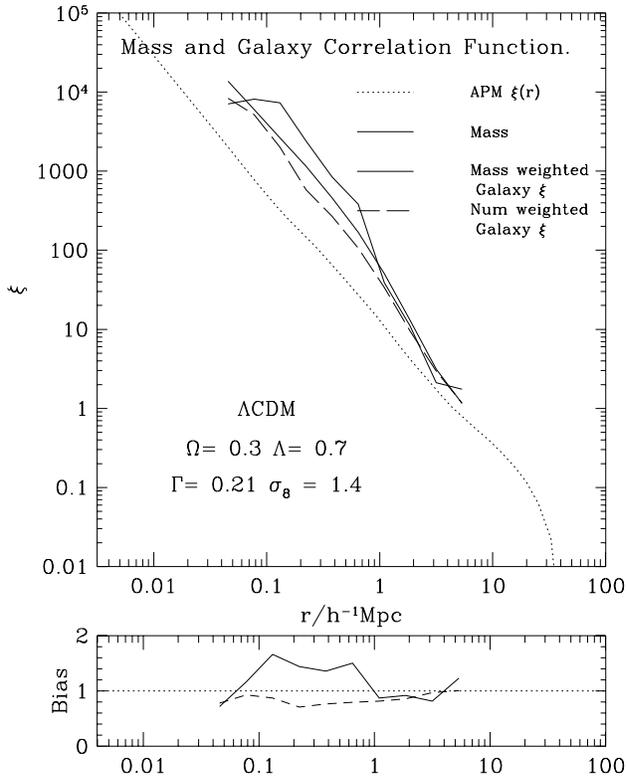}
\caption{Galaxy correlation functions in the \LCDM\
simulation. The dashed line shows the correlation function with galaxies
weighted equally and the thin solid line shows the correlation function
with each galaxy weighted in proportion to its gas mass.  The other curves
are as in Figure~3.}
\end{figure}  

The weaker clustering of the galaxy distribution compared to the mass in
the \LCDM\ model is intriguing. This is the first time that such an effect
has been seen in a cosmological simulation. A clue to its cause is provided
by comparing the (number weighted) galaxy correlation function of Figure~3
with the galaxy-mass weighted correlation function. This comparison is
carried out in Figure~4 and shows that the antibias is replaced by a
positive bias when galaxies are weighted by their mass rather than
equally. This suggests that the antibias present in the galaxy
distribution is related to the merging and disruption of galaxies in rich
clusters which produces very massive objects at their centres. We are
currently carrying out a series of tests to examine this issue in detail. 

Whilst the physics in our dark matter simulations are well understood and
the results are reliable, the same cannot yet be said of our dark matter
plus gas simulations. We are continuing the analysis of these and related
simulations in an effort to better understand the numerics and physics in a
regime not previously explored.  In particular, we stress that the galaxy
correlation functions are very sensitive to the distribution of galaxies in
the cores of rich clusters. The formation of excessively massive galaxies
in these cores suggests that firm conclusions regarding biases in the
galaxy distribution will have to await a better understanding of the
processes at work in these extreme environments.

\acknowledgments
The simulations reported here were carried out on the Cray-T3Ds at the
Edinburgh Parallel Computing Centre, and the Rechen Zentrum Garching.
This work was supported by grants from PPARC and EPSRC.


%
%

%


\begin{references}
\reference Baugh, C. M. 1996, \mnras, 280,267
\reference Bernardeau, F., 1994, \apj,433,1
\reference Bond J. R., Cole S., Efstathiou, G. and Kaiser, N.,1991,\apj,379,440
\reference Bower, R. G. 1992, \mnras, 238,332
\reference Bouchet, F., Juskiewicz R., Columbi, S. and Pellat, R.,1992,\apj 394,L5 
\reference Cen, R. and Ostriker, J.,1996,\apj,464,270
\reference Cole, S., Aragon-Salamanca, A., Frenk, C. S., Navrro, J. F.
and Zepf, S. E., 1994,\mnras,271,781
\reference Copi, C. J., Schramm, D. N., Turner, M. S., 1995,\apj,455,95
\reference Couchman, H. M. P., Thomas, P. A. \& Pearce F. R.
 1995, \apj, 452, 797
\reference Eke, V. R., Cole, S. and Frenk, C. S. 1996, \mnras, 282.263
\reference Evrard, A., Summers, F. and Davis, M., 1994, \apj,422,11
\reference Frenk, C. S., 1991,{\it Models of large scale structure}, in {Nobel 
Symposium No. 79: 
The birth and early evolution of our universe}, { Physica Scripta}, 
{\bf  T36}, 70-87
\reference Frenk, C. S., Evrard, A. E., White, S. D. M. and Summers, F. J., 
1996,\apj,in press
\reference Kauffmann, G., White, S. D. M. and Guiderdoni, B., 1993, \mnras,
264,201
\reference Katz, N., Hernquist, L. and Weinberg, D. H., 1992,\apj 399,109
\reference Maddox, S., Efstathiou G., Sutherland, W. and Loveday J.,\mnras,242,43p
\reference Pearce, F.R., Couchman, H.M.P., Jenkins, A.R., Thomas, P.A., 1995,
`Hydra - Resolving a parallel nightmare', in Dynamic Load Balancing on 
 MPP systems.
\reference Press, W. H. and Schechter, P., 1974,\apj,187,425
\reference White, S. D. M., Efstathiou, G. P. and Frenk, C. S. 1993, 
\mnras,262,1023
\reference White, S. D. M. and Frenk, C. S., 1991, \apj,379,52




\end{references}
\end{document}